\documentclass[12pt,preprint]{aastex}

\begin{document}

\title{GLOBAL SLIM ACCRETION DISK SOLUTIONS REVISITED}

\author{Cheng-Liang Jiao, Li Xue, Wei-Min Gu, and Ju-Fu Lu}
\affil{Department of Physics and Institute of Theoretical Physics
and Astrophysics, \\ Xiamen University, Xiamen, Fujian 361005,
China; lujf@xmu.edu.cn}

\begin{abstract}
We show that there exists a maximal possible accretion rate, beyond
which global slim disk solutions cannot be constructed because in
the vertical direction the gravitational force would be unable to
balance the pressure force to gather the accreted matter. The
principle for this restriction is the same as that for the Eddington
luminosity and the corresponding critical accretion rate, which were
derived for spherical accretion by considering the same force
balance in the radial direction. If the assumption of hydrostatic
equilibrium is waived and vertical motion is included, this
restriction may become even more serious as the value of the maximal
possible accretion rate becomes smaller. Previous understanding in
the literature that global slim disk solutions could stand for any
large accretion rates is due to the overestimation of the vertical
gravitational force by using an approximate potential. For accretion
flows with large accretion rates at large radii, outflows seem
unavoidable in order for the accretion flow to reduce the accretion
rate and follow a global solution till the central black hole.
\end{abstract}

\keywords{ accretion, accretion disks - black hole physics -
hydrodynamics}

\section{INTRODUCTION}
Although great progress has been made in recent years in
increasingly sophisticated numerical accretion disk simulations,
simple analytic disk models still are the only accessible way of
making direct link between the theory and observations, as only
these models can be used to estimate, e.g., the spectra of
accretion-powered astrophysical systems. This fact justifies the
continuous effort to improve the understanding of accretion
processes using a simple viscosity parameterization.

The slim disk model is one of such simple analytic models for black
hole accretion (Abramowicz et al. 1988; Kato et al. 1998). This
model was developed upon the standard Shakura-Sunyaev disk (SSD)
model (Shakura \& Sunyaev 1973) by considering two processes in
black hole accretion flows, namely the transonic motion and the
advective heat transport, which were neglected in the SSD model. In
the limit of low mass accretion rates, i.e., $\dot{M}$ is
substantially lower than its critical value $\dot{M}_\mathrm{Edd}$
corresponding to the Eddington luminosity $L_\mathrm{Edd}$, say,
$\dot{M}\lesssim 0.1\dot{M}_\mathrm{Edd}$, the advective heat
transport is unimportant, and the structure of slim disks is similar
to that of SSDs, with a difference that slim disk flows are
transonic in their inner regions (e.g., Chen \& Taam 1993). In this
sense, the theoretical basis of black hole disk models for low and
even moderate accretion rates is well established. But the slim disk
model is supposed to have the advantage over the SSD model that it
can be extended to the case of high accretion rates, i.e., with
$\dot{M}$ approaching or surpassing $\dot{M}_\mathrm{Edd}$. In this
case, the very basic assumption of the SSD model that is valid only
for low accretion rates, i.e., the geometrical thinness, $H\ll r$,
where $H$ is the half-thickness of the disk and $r$ is the
cylindrical radius, would break down (Frank et al. 2002, p. 98). It
was then suggested in the slim disk model that the disk becomes
geometrically slim, i.e., with $H\lesssim r$; and accordingly, the
process of advective heat transport becomes important or even
dominant over the radiative cooling because, as given by Abramowicz
et al. (1986), there is a relation $f_\mathrm{adv}\varpropto
(H/r)^2$, where $f_\mathrm{adv}\equiv
Q_\mathrm{adv}^-/Q_\mathrm{vis}^+$ is the advective factor, with
$Q_\mathrm{adv}^-$ and $Q_\mathrm{vis}^+$ being the advective
cooling and viscous heating rates per unit area of the disk,
respectively.

However, the self-consistency of the slim disk model and its
applicability to the high accretion rate case are not so obvious. In
particular, uncertainties seem to concentrate on the treatment of
the vertical structure of the disk. First, vertical hydrostatic
equilibrium is a reasonable assumption for SSDs because, for
geometrically thin disks, vertical motion of the disk matter must be
negligible compared with radial motion; but it is questionable to
adopt this assumption for slim disks that are not thin. Second, as
noticed recently by Gu \& Lu (2007, hereafter GL07), even the
vertical hydrostatic equilibrium is assumed, in the slim disk model
there is a serious inconsistency that the vertical gravitational
force was greatly magnified by using the approximate form of
potential due to H\=oshi (1977), which is valid only for thin disks
again; and accordingly, the previous understanding that slim disks
could exist for any large accretion rates seems doubtful. Third, the
above mentioned relation $f_\mathrm{adv}\varpropto (H/r)^2$ was
derived with the H\=oshi approximation of potential, so its
applicability to not thin disks with large accretion rates is not
justified; or in other words, it is not clear whether an accretion
disk can ensure advection dominance (i.e., $f_\mathrm{adv} > 0.5$)
while remaining to be geometrically slim. According to the analyses
of Narayan \& Yi (1995) and Gu et al. (2008), in order for advection
to be dominant, the disk needs to be geometrically thick, i.e., with
$H > r$, rather than slim. This is probably the reason why $H > r$
was obtained in many numerical calculations of slim disks and other
accretion disks (see references in GL07).

What about another popular black hole accretion disk model, the
advection-dominated accretion flow (ADAF) model (Narayan \& Yi 1994;
Abramowicz et al. 1995)? ADAFs are also supposed to be geometrically
slim and advection-dominated, and their vertical structure was
treated in a way similar to that for slim disks. The above mentioned
uncertainties should also apply to the ADAF model, because they are
based on purely hydrostatic considerations and are related only to
the geometrical thickness of the disk. Many two- and
three-dimensional numerical simulations of viscous radiatively
inefficient accretion flows revealed the existence of
convection-dominated accretion flows rather than ADAFs (see
references in GL07). This fact is probably an indication that the
ADAF model might have hidden inconsistencies, and one of which might
be related to the treatment of vertical structure. What is different
from slim disks is that ADAFs are expected to correspond to very low
accretion rates and are known to have a maximal possible accretion
rate at each radius (e.g., Abramowicz et al. 1995), so the problem
addressed in GL07 regarding the allowed accretion rate may have no
impact on ADAFs. In addition, ADAFs are optically thin and ion
pressure-supported, the radiation processes in them are more
complicated, making the vertical structure more difficult to deal
with, than for slim disks that are optically thick and radiation
pressure-supported.

Our present work is devoted to discuss the slim disk model and is a
straightforward continuation of GL07. All the results of GL07 were
based on a local analysis, i.e., only for a certain radius. Although
a similar local analysis was often used in the literature (e.g.,
Abramowicz et al. 1995; Chen et al. 1995; Kato et al. 1998), one
should be cautious of the fact that conclusions made in the local
sense do not necessarily hold in the global sense. For example, it
has been shown that, even though a disk is locally unstable at a
given radius according to a local stability analysis, the disk can
be globally stable in global numerical simulations (Janiuk et al.
2002; Gierli\'{n}ski \& Done 2004). Therefore, it is worthwhile to
check and extend the results of GL07 by investigating global
solutions of original differential equations for black hole
accretion flows, similar to what was done by, e.g., Chen \& Wang
(2004), Watarai et al. (2005), Artemova et al. (2006), and Watarai
(2006), but with a revised vertical gravitational force.

\section{SOLUTIONS WITH VERTICAL HYDROSTATIC EQUILIBRIUM}
\subsection{Equations}
 The basic equations to be solved for slim disks can be written in cylindrical coordinates
  as (cf. Kato et al. 1998, p. 236; GL07):
\begin{equation}\label{1}
    \dot{M}=-2\pi r\Sigma v_r={\mathrm{constant}},
\end{equation}

\begin{equation}\label{2}
    v_r\frac{dv_r}{dr}+\frac{1}{\rho_0}\frac{dp_0}{dr}+(\Omega_{\mathrm{K}}^2-\Omega^2)r=0,
\end{equation}

\begin{equation}\label{3}
\dot{M}(\Omega r^2-j)=2\pi\alpha r^2\Pi,
\end{equation}

\begin{equation}\label{4}
Q_{\mathrm{vis}}^{+}=Q_{\mathrm{adv}}^{-}+Q_{\mathrm{rad}}^{-},
\end{equation}

\begin{equation}\label{5}
    p_0=\frac{k_\mathrm{B}\rho_0 T_0}{\mu m_p}+\frac{1}{3}aT^4_0,
\end{equation}
where $\Sigma=2\int_0^H\rho dz$ is the surface density, $\rho$ is
the density, $p$ is the pressure, $v_r$ is the radial velocity,
$\Omega$ is the angular velocity and $\Omega_\mathrm{K}$ is its
Keplerian value, $j$ is an integration constant representing the
specific angular momentum accreted by the black hole, $\alpha$ is
the Shakura-Sunyaev viscosity parameter, $\Pi=2\int_0^H p dz$ is the
vertically integrated pressure, $Q_\mathrm{rad}^-$ is the radiative
cooling rate per unit area, $T$ is the temperature, $\mu$ is the
mean molecular weight and is taken to be 0.62, and the subscript
``0" represents quantities on the equatorial plane.

There are some differences between equations (1 - 5) and the basic
equations in GL07 (their eqs. [6 - 10]). In GL07 all the five
equations were written in the vertically integrated form; here the
continuity equations (1), angular momentum equation (3), and energy
equation (4) are in the same form, because it is obviously
convenient to write equation (1) in this form and equations (3) and
(4) also contain $\dot{M}$; but the radial momentum equation (2) and
state equation (5) are given for equatorial quantities. The state
equation of GL07 (their eq. [10]) is a trivial vertical integration
of our equation (5) and makes no difference. But in GL07 (and also
in Kato et al. 1998), in obtaining the vertically integrated radial
momentum equation (eq. [7] of GL07 or eq. [8.25] of Kato et al.
1998), some simplifications were made. First, both $v_r$ and
$\Omega$ were regarded to be independent of the coordinate $z$, and
this is also adopted in our equations here. Second, a step
$\int_{-H}^H\frac{dp}{dr} dz=\frac{d\Pi}{dr}$ was taken, but in the
strict sense this equality holds only when $H$ does not vary with
$r$, in studies of global solutions here we should better not to
take it. This is the reason why we prefer to write equation (2) for
equatorial quantities. The differential equation (4) was reduced to
be an algebraic one in GL07; and here instead, it keeps its original
form with the explicit expressions for $Q_\mathrm{vis}^+$,
$Q_\mathrm{adv}^-$, and $Q_\mathrm{rad}^-$ given as

\begin{equation}\label{6}
Q_{\mathrm{vis}}^{+}=-\frac{\dot{M}\Omega(\Omega r^2-j)}{2\pi
r^2}\frac{d\ln{\Omega}}{d\ln{r}},
\end{equation}

\begin{equation}\label{7}
Q_{\mathrm{adv}}^{-}=-\frac{3\dot{M}}{2\pi
r^2}\frac{\Pi}{\Sigma}\left(\frac{d\ln p_0}{d\ln
r}-\frac{4}{3}\frac{d\ln\rho_0}{d\ln r}\right),
\end{equation}

\begin{equation}\label{8}
Q_{\mathrm{rad}}^{-}=\frac{32\sigma T_0^4}{3\overline{\kappa}\rho_0
H}
\end{equation}
(Kato et al. 1998), where $\overline{\kappa}=\kappa_{{\mathrm{es}}}
+ \kappa_{{\mathrm{ff}}} = 0.34 +
6.4\times{10^{22}}\overline{\rho}\overline{T}^{\ -7/2} \mathrm{cm^2\
g^{-1}}$, $\overline{\rho}(=\Sigma/2H)$ and $\overline{T}(=\int_0^H
Tdz/H)$ are the vertically averaged density and temperature,
respectively.

The key difference between the existing slim disk model and our work
here is in the treatment of the vertical hydrostatic equilibrium
equation,

\begin{equation}\label{9}
\frac{\partial p}{\partial z}+\rho \frac{\partial\psi}{\partial
z}=0,
\end{equation}
where $\psi$ is the potential. In the slim disk model (e.g., Kato et
al. 1998, p. 241), the potential of Paczy\'nski \& Wiita (1980),

\begin{equation}\label{10}
\psi(r,z)=-\frac{GM}{\sqrt{r^2+z^2}-r_g},
\end{equation}
where $r_g\equiv2GM/c^2$ is the gravitational radius, was
approximated in the form of H\=oshi (1977), i.e.,

\begin{equation}\label{11}
\psi(r,z)\approx\psi(r,0)+\frac{\Omega_{\mathrm{K}}^2 z^2}{2}.
\end{equation}
Using equation (11) and assuming a polytropic relation in the
vertical direction, $p=K\rho^{1+1/N}$, where $K$ and $N$ are
constants, the vertical integration of equation (9) gave (for $N =
3$)

\begin{equation}\label{12}
    \left(\frac{\rho}{\rho_0}\right)^{1/3}=\left(\frac{p}{p_0}\right)^{1/4}=\frac{T}{T_0}=1-\frac{z^2}{H^2},
\end{equation}

\begin{equation}\label{13}
    8\frac{p_0}{\rho_0}=\Omega_{\mathrm{K}}^2H^2,
\end{equation}
and $\Sigma$ and $\Pi$ were

\begin{equation}\label{14}
    \Sigma=2\times \frac{16}{35}\rho_0H,
\end{equation}

\begin{equation}\label{15}
    \Pi=2\times\frac{128}{315}p_0H.
\end{equation}
The simple relation ${c_s}/\Omega_{\mathrm{K}}H=\mathrm{constant}$
was obtained, with the sound speed $c_s$ being defined either as
$c_s^2=p_0/\rho_0$ or as $c_s^2=\Pi/\Sigma$.

However, as the main point made in GL07, the approximation of
equation (11) is invalid for slim disks. By using the explicit
Paczy\'nski \& Wiita potential, equation (10), to integrate equation
(9), one obtains instead of equations (12) and (13):

\begin{equation}\label{16}
    g=\left(\frac{\rho}{\rho_0}\right)^{1/3}=\left(\frac{p}{p_0}\right)^{1/4}
    =\frac{T}{T_0}=\frac{\frac{1}{\sqrt{r^2+z^2}-r_g}-\frac{1}{\sqrt{r^2+H^2}-r_g}}
    {\frac{1}{r-r_g}-\frac{1}{\sqrt{r^2+H^2}-r_g}},
\end{equation}

\begin{equation}\label{17}
    4\frac{p_0}{\rho_0}=GM\left(
    \frac{1}{r-r_g}-\frac{1}{\sqrt{r^2+H^2}-r_g}\right).
\end{equation}
Accordingly, $\Sigma$ and $\Pi$ become

\begin{equation}\label{18}
    \Sigma=2\rho_0\int_0^Hg^3dz,
\end{equation}

\begin{equation}\label{19}
    \Pi=2p_0\int_0^Hg^4dz,
\end{equation}
which are much more complicated than equations (14) and (15). It is
seen that the relation
${c_s}/\Omega_{\mathrm{K}}H=\mathrm{constant}$ does not hold.

To summarize, in the existing slim disk model, the set of eight
equations, namely equations (1 - 5) and (13 - 15), could be solved
for the eight unknown quantities $\rho_0$, $p_0$, $\Sigma$, $\Pi$,
$T_0$, $v_r$, $\Omega$, and $H$ as functions of $r$, with given
constant parameters $M$, $\dot{M}$, $\alpha$, and $j$; while in our
work here, equations (17 - 19), instead of equations (13 - 15),
still along with equations (1 - 5) form a new set to be solved for
the same eight unknowns. There are two differential equations, i.e.,
equations (2) and (4); and other equations in each set are
algebraic, including equations (18) and (19) once the vertical
integration is analytically made.

\subsection{Solutions}

We numerically solve the set of equations (1 - 5) and (17 - 19),
i.e., with the correctly calculated vertical gravitational force
using the explicit Paczy\'nski \& Wiita potential. For comparisons,
we also solve the set of equations (1 - 5) and (13 - 15), as done
already in the existing slim disk model, i.e., with the magnified
vertical gravitational force using the H\=oshi approximation of the
Paczy\'nski \& Wiita potential. After some algebra, the equations in
each set can be combined into only two differential equations. A
physically acceptable global solution should be able to extend from
a large radius to the vicinity of the central black hole, passing
the sonic point regularly. We start the integration at $r =
10^6r_g$, where the outer boundary conditions are set to be
corresponding to an SSD, i.e., being geometrically thin and
Keplerian rotating. If a solution is found to be transonic, we
extend it to $r = 2r_g$ (in the following figures only a smaller
radial range is shown in order to see more clearly the solution
behavior in the inner region). We fix $M = 10M_\sun$ and $\alpha =
0.1$, and vary the values of $\dot{M}$ for different solutions. For
a transonic solution, the other constant parameter $j$ is an
eigenvalue of the problem and has to be adjusted correctly and
accurately (e.g., Chen \& Taam 1993; Chen \& Wang 2004).

Rather than showing the radial variation of every physical quantity,
our main purpose here is to see how the correction of the vertical
gravitational force changes the previous understanding of global
slim disk solutions, that is, there was no upper limit of $\dot{M}$
for slim disks, any large value of $\dot{M}$ could correspond to a
global solution (e.g., Kato et al. 1998; Chen \& Wang 2004; Watarai
et al. 2005; Watarai 2006). Figures 1 and 2 are sufficient for this
purpose, which are for the relative thickness $H/r$ and the
advective factor $f_\mathrm{adv}$, respectively. In the figures,
solid lines are solutions of equations (1 - 5) and (17 - 19), and
dashed lines are solutions of equations (1 - 5) and (13 - 15). Solid
line $a$ and dashed line $a$' are for the same accretion rate
$\dot{m}= 1$, where $\dot{m}\equiv\dot{M}/\dot{M}_\mathrm{Edd}$,
with $\dot{M}_\mathrm{Edd} = 64\pi GM/c\kappa_\mathrm{es}$ being the
Eddington accretion rate; and similarly, solid line $b$ and dashed
line $b$', solid line $c$ and dashed line $c$', solid line $d$ and
dashed line $d$', and solid line $e$ and dashed line $e$' are for
$\dot{m}= 10, 31.5, 31.7$, and 100, respectively. It is seen that
for a moderate accretion rate $\dot{m}= 1$ (lines $a$ and $a$'),
global solutions obtained from the two sets of equations are almost
the same, i.e., both the solutions have $H/r\lesssim 1$ (Fig. 1),
and are radiation-dominated in the outer region and with important
advection in the inner region (Fig. 2). This proves that using the
H\=oshi approximation of potential for moderate accretion rates is
acceptable. However, deviations appear and become serious as
$\dot{m}$ increases. For $\dot{m}= 10$, our new solution (lines $b$
in the two figures) has $H$ and $f_\mathrm{adv}$ significantly
larger than that in the existing slim disk solution (lines $b$').
Such a situation continues till a critical value $\dot{m}= 31.5$,
for which our new solution can still be constructed (lines $c$). For
a slightly larger $\dot{m}= 31.7$ and a still larger $\dot{m}= 100$,
global solutions can be no longer obtained from our new set of
equations (1 - 5) and (17 - 19). It is seen that $H$ tends to
infinity (lines $d$ and $e$ in Fig. 1) and $f_\mathrm{adv}$ tends to
exceed 1 (its maximal possible value, lines $d$ and $e$ in Fig. 2),
so the inward integration cannot go on. These solutions (lines $d$
and $e$) are not global solutions at all, since they cannot extend
to a sonic point. On the other hand, no matter how large $\dot{M}$
is, global solutions can always be found from the set of existing
slim disk equations (1 - 5) and (13 - 15), as drawn by lines $b$',
$c$', $d$', and $e$' in the two figures (lines $c$' and $d$'
coincide with each other).

Unfortunately, the existing global slim disk solutions, though being
formally constructed for any high accretion rates, have hidden
inconsistencies. As seen clearly in Figure 1 of GL07, the vertical
gravitational force, $\partial\psi/\partial z$ in equation (9), was
greatly overestimated, and accordingly, the geometrical thickness
$H$ was greatly underestimated, by using the approximation of
equation (11). This is the reason why in the slim disk model, the
gravitational force seemed to be always able to balance the pressure
force and ensure vertical hydrostatic equilibrium. Even so, slim
disk solutions may still have $H/r > 1$ (lines $c$', $d$', and $e$'
in Fig. 1), the assumption of slimness is violated.

As the main result of our work, it is found that, when the vertical
gravitational force is correctly calculated from equation (10),
there exists a maximal possible accretion rate,
$\dot{M}_\mathrm{max}\approx 31.5\dot{M}_\mathrm{Edd}$, beyond which
there are no global solutions at all (lines $d$ and $e$ in Figs. 1
and 2). The physical reason for this is the following. The amount of
accreted matter that can be gathered by the black hole's
gravitational force must be limited. If $\dot{M} >
\dot{M}_\mathrm{max}$, the pressure force of the matter and
radiation in the vertical direction would be too large to be
balanced by the gravitational force, the disk would be huffed by the
pressure force to tend to an infinite thickness (lines $d$ and $e$
in Fig. 1), and the accretion processes would never be maintained.
This reason is based on the same principle as that for the Eddington
luminosity $L_\mathrm{Edd}$ and the corresponding critical accretion
rate $\dot{M}_\mathrm{Edd}$ to be defined. The only difference is
that $L_\mathrm{Edd}$ and $\dot{M}_\mathrm{Edd}$ were derived for
spherical accretion, i.e., by considering the balance between the
gravitational force and the pressure force in the radial direction;
and here the maximal possible accretion rate is found by considering
the balance of the same two forces in the vertical direction of
accretion disks.

In GL07, by a local analysis, i.e., considering the vertical balance
of gravitational and pressure forces at a certain radius, a similar
maximal possible accretion rate, $\dot{M}_\mathrm{max}(r)$, was
found for each radius. Their result is confirmed here. But a global
solution is with a constant accretion rate, so the allowed accretion
rate has a unique value, rather than being radius-dependent.

As a check of our global solutions, the total optical depth
$\tau=\overline{\kappa} \Sigma/2$ is calculated and is shown in
Figure 3, where lines $a$, $b$, $c$, $a$', $b$', and $c$' are the
correspondents of lines $a$, $b$, $c$, $a$', $b$', and $c$' in
Figures 1 and 2, respectively. It is seen that, similar to the
existing slim disk solutions (lines $a$', $b$', and $c$'), our
global solutions are also optically thick everywhere (lines $a$,
$b$, and $c$). But $\tau$ in our solutions is somewhat smaller than
that in the existing slim disk solutions, especially in the inner
regions. This is because in our solutions, the correctly calculated
vertical gravitational force is smaller, then $H$ is larger, $v_r$
is larger, and $\Sigma$ is smaller for the same $\dot{M}$; and
$\overline{\kappa}$ is almost unchanged, especially in the inner
regions where the electron scattering opacity is dominant and is a
constant.

\section{DISCUSSION}

\subsection{Waiving hydrostatic equilibrium}

Though global solutions with vertical gravitational force correctly
calculated can be obtained for $\dot{M} < \dot{M}_\mathrm{max}$ as
shown in Figures 1, 2, and 3, there seems to be also some
inconsistency hidden in these solutions. For high accretion rates,
the disk's relative thickness $H/r$ becomes substantially larger
than 1 (lines $b$ and $c$ in Fig. 1); and for the critical value
$\dot{M}=31.5 \dot{M}_\mathrm{Edd}$, $H/r$ reaches up to $\sim20$ in
the middle region of the solution. As argued by Abramowicz et al.
(1997), if it is assumed that there is no velocity component in the
direction orthogonal to the surface of the disk (i.e., no outflows
leaving the disk), then there is a relation $v_H/v_r=dH/dr$, where
$v_H$ is the vertical velocity on the surface, and vertical motion
should not be neglected if $H$ does not very slowly vary with $r$.
In particular, at the maximum of $H/r$ (see lines $b$ and $c$ in
Fig. 1), it follows from $d(H/r)/dr = 0$ that $dH/dr = H/r$. This
means that in the middle region of the solution the vertical
velocity could greatly exceed the radial velocity, and the
assumption of hydrostatic equilibrium would not be valid. In this
case, instead of hydrostatic equilibrium equation (9), one should
use the more general form of vertical momentum equation,

\begin{equation}\label{20}
\frac{1}{\rho}\frac{\partial p}{\partial
z}+\frac{\partial\psi}{\partial z}+v_r \frac{\partial v_z}{\partial
r}+v_z \frac{\partial v_z}{\partial z}=0
\end{equation}
(Abramowicz et al. 1997).

Solving this partial differential equation is beyond the capacity of
the slim disk model that is one-dimensional. In an illustrative
sense, here we wish to try to consider the non-negligible vertical
velocity $v_z$ with a very simple treatment, which is similar to
what was done in Abramowicz et al. (1997). We assume

\begin{equation}
    v_z(r,z)=\frac{z}{H}v_H=\frac{z}{r}u,
\end{equation}
where

\begin{equation}
u(r)=v_r\frac{d\mathrm{ln}H}{d\mathrm{ln}r},
\end{equation}
then equation (20) is reduced to be

\begin{equation}
\frac{1}{\rho}\frac{\partial p}{\partial
z}+\frac{\partial\psi}{\partial z}+zY=0,
\end{equation}
where

\begin{equation}
   Y(r)=\frac{1}{r^2}\left(r v_r \frac{d u}{d r}-v_r u+u^2\right),
\end{equation}
which looks similar to equation (9) of Abramowicz et al. (1997).

Instead of equations (16) and (17), the vertical integration of
equation (23) gives (of course, still with the explicit potential
eq. [10])

\begin{equation}
    g=\left(\frac{\rho}{\rho_0}\right)^{1/3}=\left(\frac{p}{p_0}\right)^{1/4}=\frac{T}{T_0}
    =\frac{G M(\frac{1}{\sqrt{r^2+z^2}-r_g}-\frac{1}{\sqrt{r^2+H^2}-r_g})+\frac{1}{2} Y(H^2-z^2)}
    {G M(\frac{1}{r-r_g}-\frac{1}{\sqrt{r^2+H^2}-r_g})+\frac{1}{2}Y H^2},
\end{equation}

\begin{equation}
    4\frac{p_0}{\rho_0}=G M\left(\frac{1}{r-r_g}-\frac{1}{\sqrt{r^2+H^2}-r_g}\right)+\frac{1}{2} Y
    H^2.
\end{equation}
Equations (18) and (19) are formally unchanged, but the quantity g
in these two equations is given by equation (25) now.

The nine equations, i.e., equations (1 - 5), (18), (19), (22), and
(26) can be solved for nine unknowns, $\rho_0$, $p_0$, $\Sigma$,
$\Pi$, $T_0$, $v_r$, $\Omega$, $H$ and $u$ (eq. [1] keeps unchanged
even there is a non-zero $v_z$, because of the no outflow
assumption; and eqs. [2 - 5] still hold as well). The situation of
solutions is seen in Figure 4 that shows $H/r$ as a function of $r$.
Global solutions exist till a new maximal possible accretion rate
$\dot{M}_\mathrm{max}\approx8.5 \dot{M}_\mathrm{Edd}$ (lines $a$ and
$b$ that are for $\dot{m} = 1$ and 8.5, respectively), above which
there are no global solutions (lines $c$ and $d$ that are for
$\dot{m} = 9$ and 30, respectively). But this time, for $\dot{M} >
\dot{M}_\mathrm{max}$, it is not the case of Figure 1 that the
inward integration stops at a radius where the thickness $H$ tends
to infinity; rather, it is the case that the inward integration
stops at a radius where the quantity $g$ of equation (25) tends to
become unphysically negative at some height $z$ between the
equatorial plane $z = 0$ (where $g = 1$) and the disk surface $z =
H$ (where $g = 0$), even though $H$ is finite at that radius. It is
not surprising that the value of $\dot{M}_\mathrm{max}$ here is even
smaller than that in the above section where vertical hydrostatic
equilibrium is assumed. The reason for this is the following. As
seen from Figure 5, in global solutions $a$ and $b$, $v_H$ is always
negative, i.e., the vertical motion is always inward towards the
equatorial plane; and the absolute value of $v_H$ increases with
decreasing $r$. This means clearly that there is a vertical
acceleration towards the equatorial plane. That is, in the vertical
direction, the gravitational force has to overcome the pressure
force to accelerate the accreted matter, rather than only balancing
the pressure force to ensure hydrostatic equilibrium. The upper
limit of the amount of the accreted matter, to which the
gravitational force is able to do this harder job, must be smaller
than in the hydrostatic equilibrium case. For $\dot{M} >
\dot{M}_\mathrm{max}$, the gravitational force would be unable to do
the job. To keep equation (23) formally holding, $g$ (and $\rho$ and
$T$) of equation (25) would have to become mathematically negative,
which are physically unacceptable.

Though the treatment of vertical motion made here is crude, it is
sayable that inclusion of this motion should not change the main
conclusion reached in the above section, i.e., there is a maximal
possible accretion rate $\dot{M}_\mathrm{max}$ for global slim disk
solutions to exist; and the value of $\dot{M}_\mathrm{max}$ becomes
even smaller than in the case that this motion was omitted.

\subsection{Outflows}

If the mass accretion rate given at large radii exceeds its maximal
possible value, $\dot{M} > \dot{M}_\mathrm{max}$, it seems that the
only possible way for a global slim disk solution to be realized is
that the accretion flow loses its matter at large radii in the form
of outflows, such that at smaller radii $\dot{M}$ is reduced to be
below $\dot{M}_\mathrm{max}$.

Outflows have been observed in many high energy astrophysical
systems that are believed to be powered by black hole accretion, but
the mechanism of outflow formation remains unclear from the
theoretical point of view. At the same time when the ADAF model was
proposed, it was suggested that ADAFs are likely to be able to
produce outflows because they have a positive Bernoulli constant in
their self-similar solutions (Narayan \& Yi 1994). Later, Abramowicz
et al. (2000) showed that in global solutions of ADAFs the Bernoulli
function, rather than the Bernoulli constant, can have either
positive or negative values, and that even a positive Bernoulli
function is only a necessary, not a sufficient, condition for
outflow formation.

Historically, the Bernoulli constant or function was defined as the
sum of the specific enthalpy, kinetic energy, and gravitational
potential energy. For ADAFs Narayan \& Yi (1994) and Abramowicz et
al. (2000) discussed, the Bernoulli function has a simple expression
since the enthalpy is totally due to the disk gas; it has also a
clear physical meaning that its positivity may imply a possibility
of outflows. For slim disks, however, the contribution from
radiation to the enthalpy becomes important or even dominant over
that from gas. To our knowledge, in this case it is unclear whether
and how can the Bernoulli function be defined, or whether this
function would have the same meaning as for ADAFs if its historical
definition is copied. Therefore, we use instead the specific total
energy of the matter in slim disks to consider the possibility of
outflows, whose equatorial-plane value is

\begin{equation}\label{27}
E=\left(\frac{3}{2}\frac{k_\mathrm{B}T_0}{\mu m_p}+\frac{a
T_0^4}{\rho_0}\right)+\frac{1}{2}\left(
v_r^2+\Omega^2r^2\right)-\frac{GM}{r-r_g}
\end{equation}
(cf. eq. [11.33] of Kato et al. 1998).

Figure 6 shows $E$ corresponding to the solutions with the correct
gravitational force in Figures 1 and 2. Lines $a$, $b$, $c$, $d$,
and $e$ in Figure 6 are for the same physical parameters as lines
$a$, $b$, $c$, $d$, and $e$ in Figures 1 and 2, respectively. For
accretion rates that allow global solutions to exist, $\dot{M}= 1,
10$ and $31.5\dot{M}_\mathrm{Edd}$ (lines $a$, $b$, and $c$), $E$ is
negative everywhere in the disk, and outflows are unlikely to
originate. However, for accretion rates exceeding
$\dot{M}_\mathrm{max}(r)$, $\dot{M}= 31.7$ and
$100\dot{M}_\mathrm{Edd}$ (lines $d$ and $e$), it is seen that as
$r$ decreases, $E$ tends to become positive first, and then the
solution stops extending. Further, it is noticeable that, when a
flow changes from a state represented by line $e$ to a state
represented by line $d$, its $E$ increases, and at the same time its
$\dot{M}$ decreases. The increase of $E$ favors outflow formation,
and the decrease of $\dot{M}$ is just the result of outflows. Such a
process of the change of the flow's state continues until $\dot{M}$
is sufficiently reduced and the flow reaches a state represented by
line $c$, then $E$ becomes negative and outflows cease.

To summarize, in this subsection we make two arguments for outflows
produced in an accretion flow with large $\dot{M}$ at large radii.
First, outflows are necessary because the accretion flow must lose
its matter in this way in order to follow a global solution till the
central black hole. Second, outflows are possible because the
accretion flow can have a positive specific total energy; what is
additionally needed for outflows to be realized is that the matter
in the accretion flow comes under an outward perturbation in the
vertical direction.

\acknowledgments

We thank Lin-Hong Chen and Jian-Min Wang for providing us their
numerical code of global slim disk solutions and the referee for
very helpful comments. This work was supported by the National Basic
Research Program of China under Grant No. 2009CB824800 and the
National Natural Science Foundation of China under Grants No.
10503003, 10673009 and 10833002.

\clearpage
\begin{figure}
\plotone{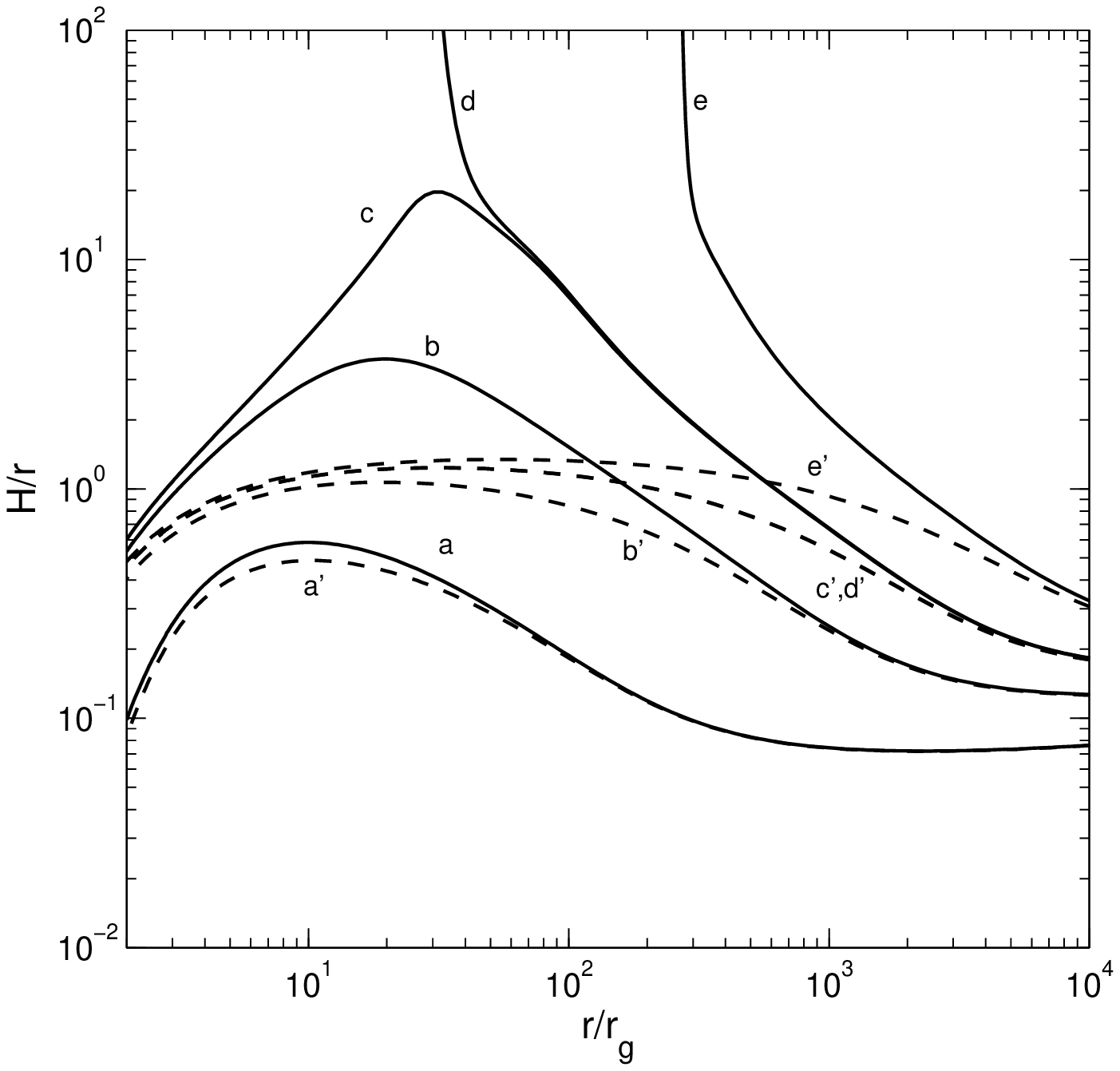} \caption{Disk's relative thickness $H/r$ as a
function of $r$. Solid lines represent our results with the vertical
gravitational force revised, and dashed lines the results in the
existing slim disk model. Lines $a$ and $a$', $b$ and $b$', $c$ and
$c$', $d$ and $d$', and $e$ and $e$' are for accretion rates
$\dot{M}/\dot{M}_\mathrm{Edd}=$ 1, 10, 31.5, 31.7, and 100,
respectively. There exists a maximal possible accretion rate
$\dot{M}_\mathrm{max}\approx 31.5\dot{M}_\mathrm{Edd}$, above which
global solutions cannot be constructed as $H$ tends to infinity
(lines $d$ and $e$). Here and in Figs. 2, 3, and 6 the vertical
hydrostatic equilibrium is assumed.} \label{fig1}
\end{figure}

\begin{figure}
\plotone{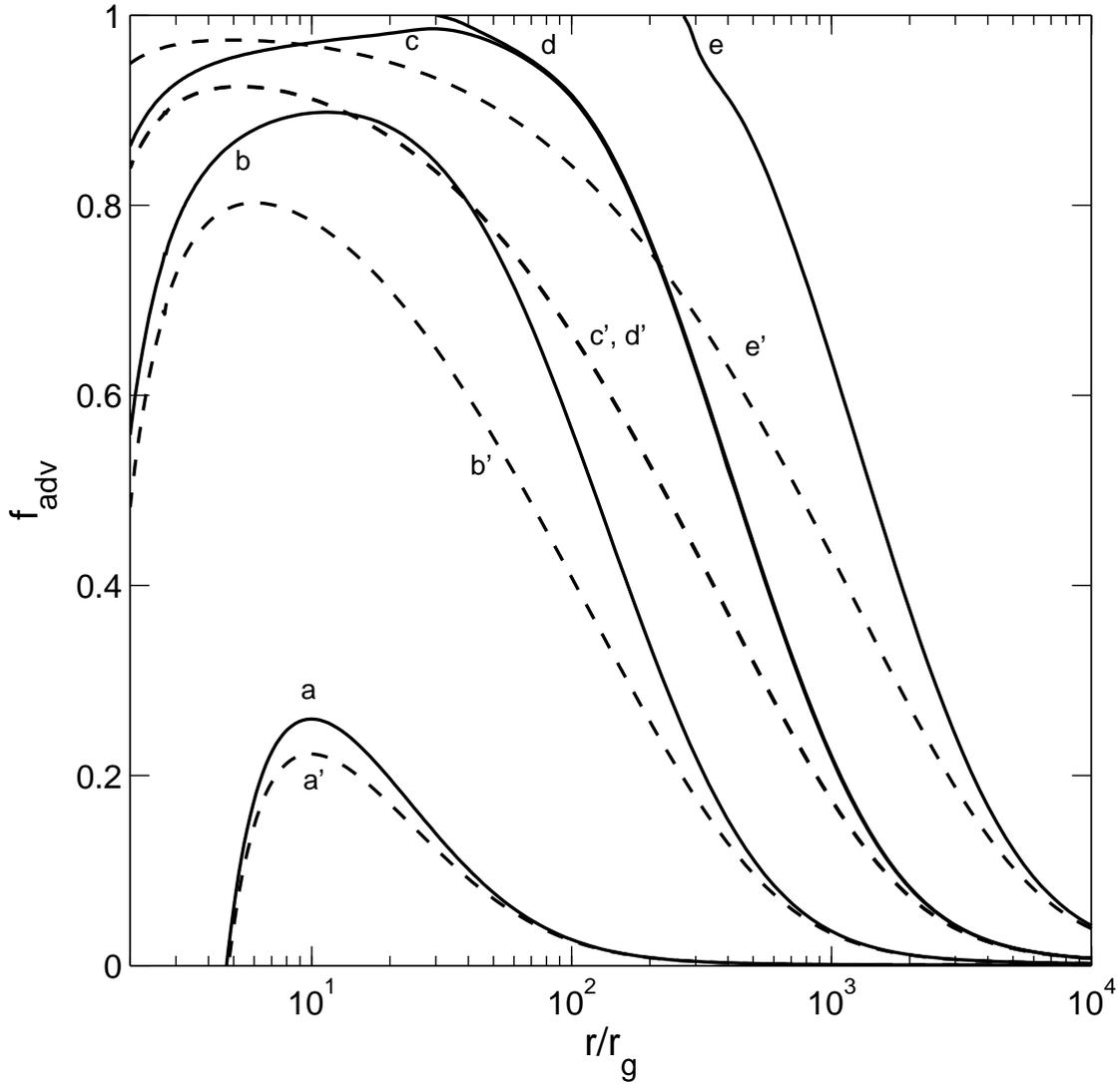} \caption{Advective factor $f_\mathrm{adv}$ as a
function of $r$. All the lines are the correspondents of those in
Fig. 1, respectively. Lines $d$ and $e$ are not global solutions as
$f_\mathrm{adv}$ tends to exceed its maximal possible value 1.}
\label{fig2}
\end{figure}

\begin{figure}
\plotone{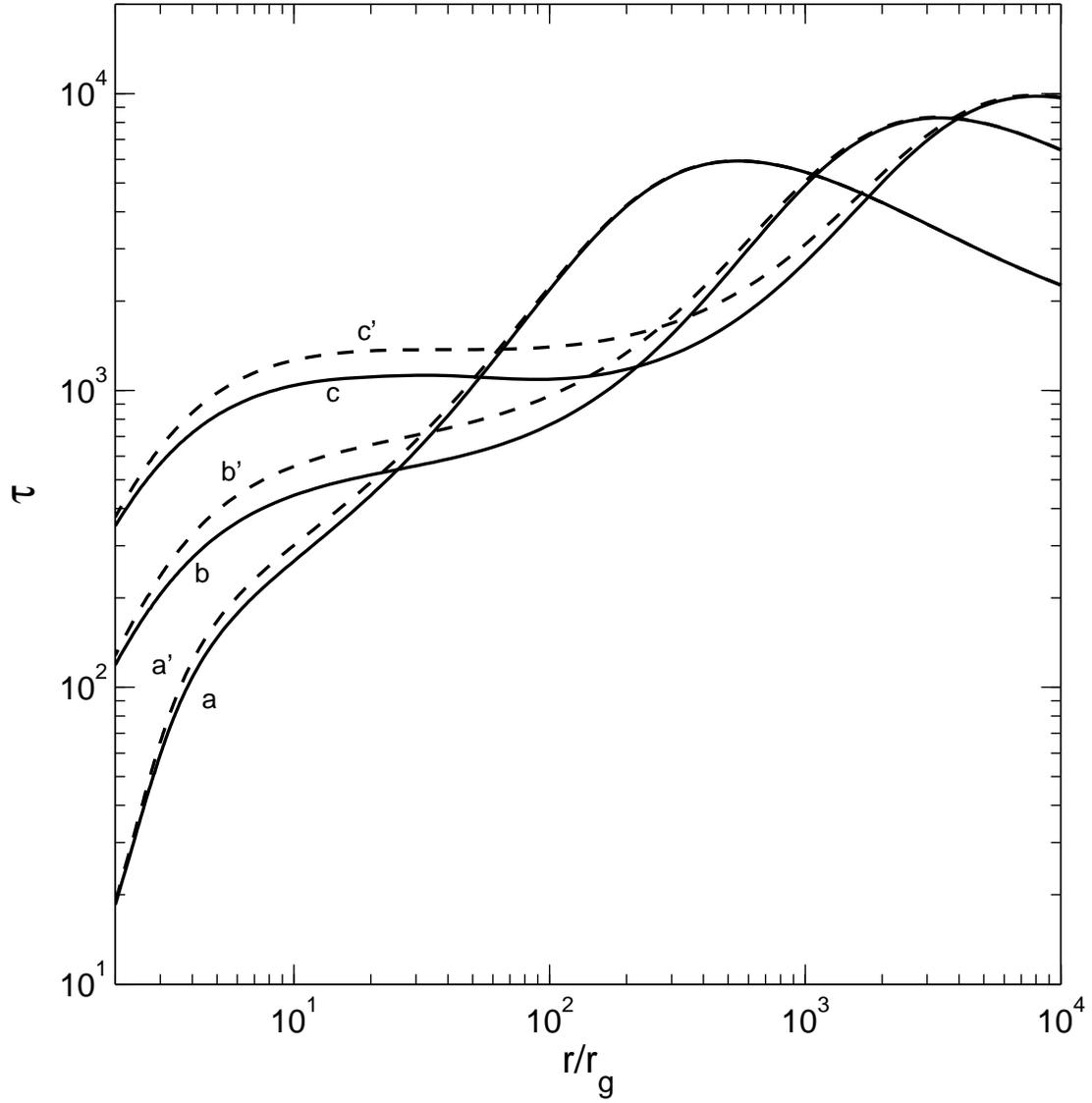} \caption{Total optical depth $\tau$ as a function
of $r$. Lines $a$, $b$, $c$, $a$', $b$', and $c$' are the
correspondents of those in Fig. 1, respectively.} \label{fig3}
\end{figure}

\begin{figure}
\plotone{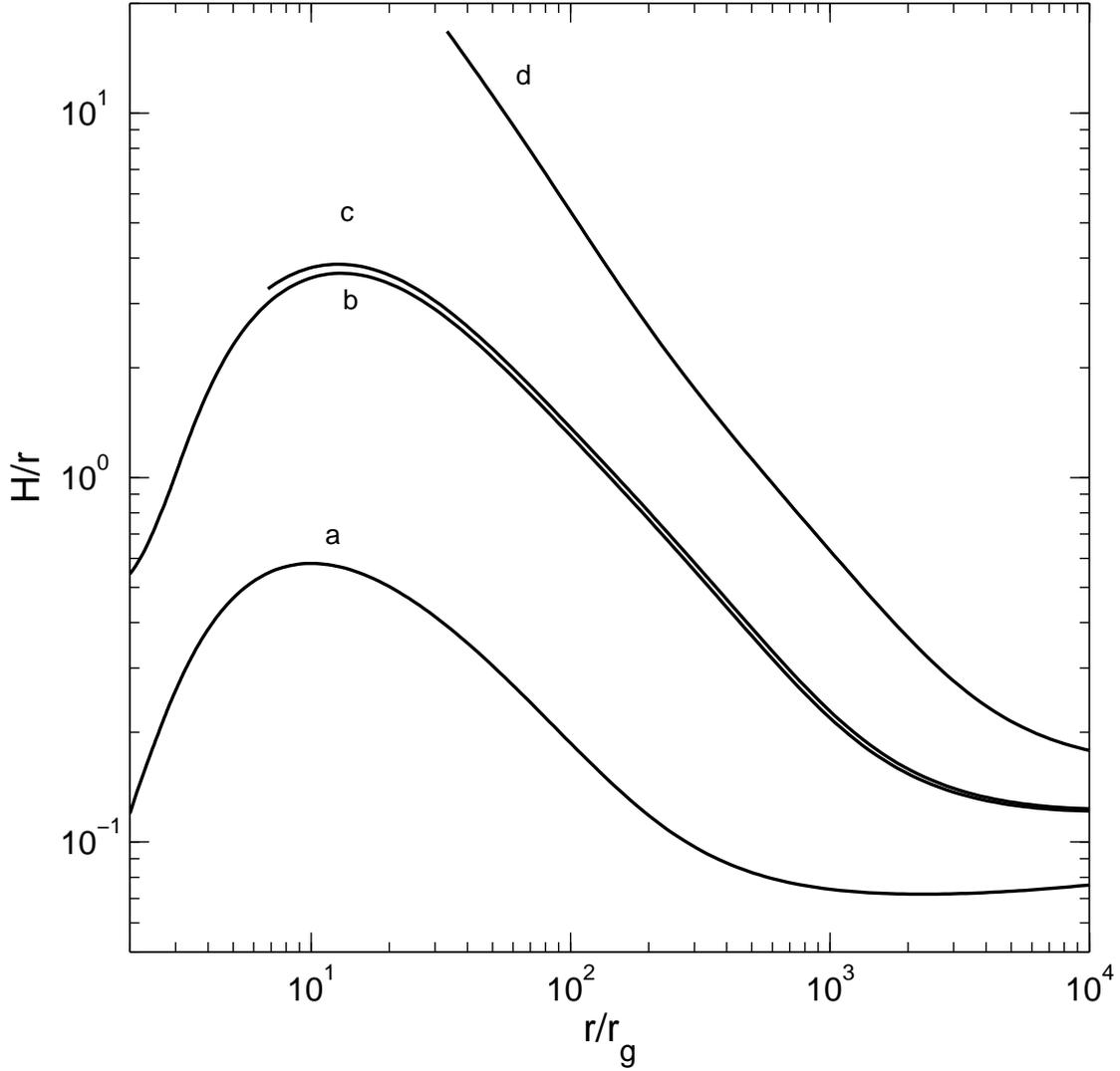} \caption{ Quantity $H/r$ as a function of $r$ in
the case that vertical motion is included. Lines $a$, $b$, $c$, and
$d$ are for $\dot{M}/\dot{M}_\mathrm{Edd}=$ 1, 8.5, 9, and 30,
respectively. There is a new $\dot{M}_\mathrm{max}\approx
8.5\dot{M}_\mathrm{Edd}$, above which there are no global solutions
(lines $c$ and $d$). }\label{fig4}
\end{figure}

\begin{figure}
\plotone{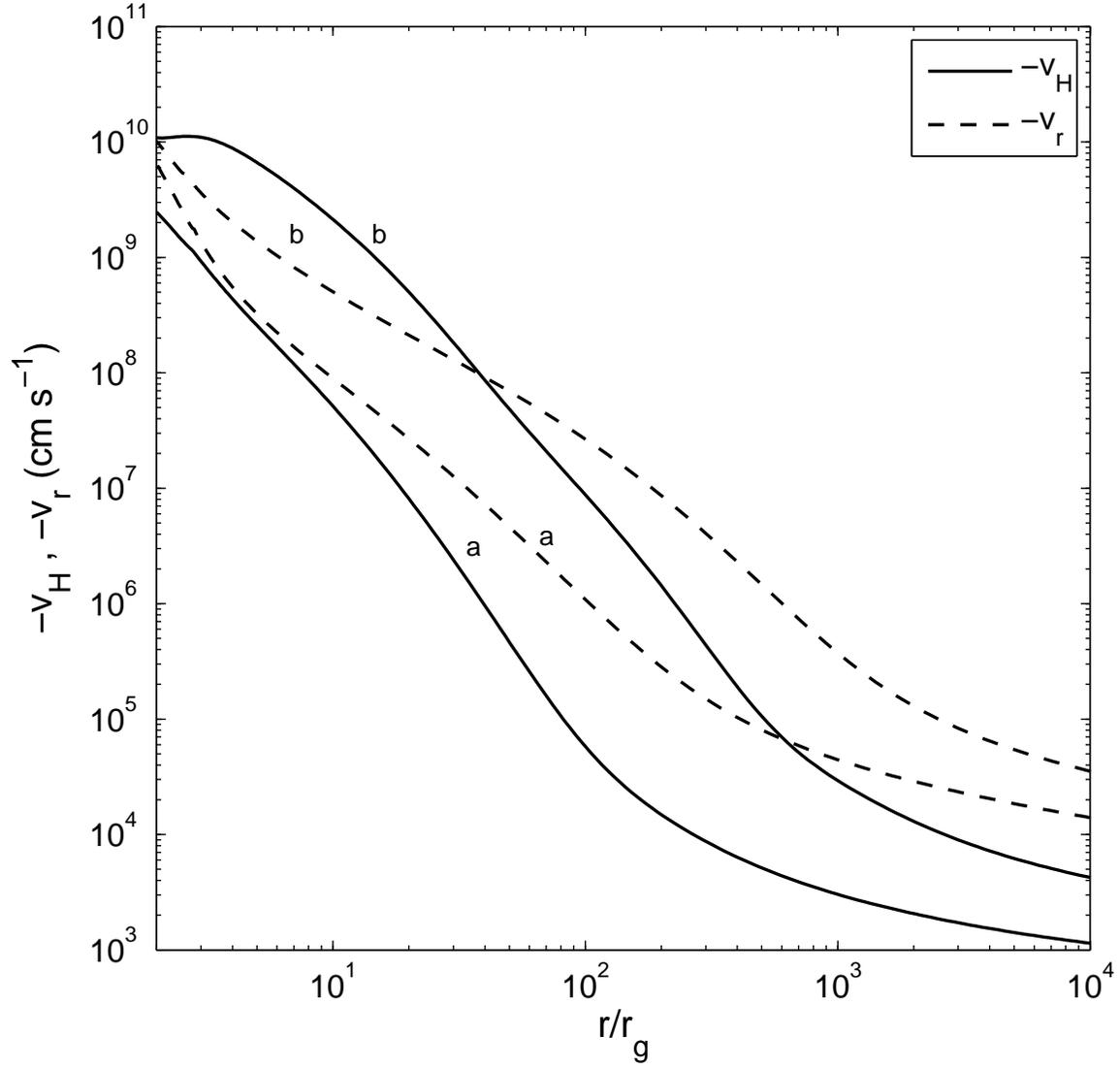} \caption{Radial velocity $v_r$ and vertical
velocity on the disk surface $v_H$ in global solutions. Lines $a$
and $b$ are the correspondents of those in Fig. 4,
respectively.}\label{fig5}
\end{figure}

\begin{figure}
\plotone{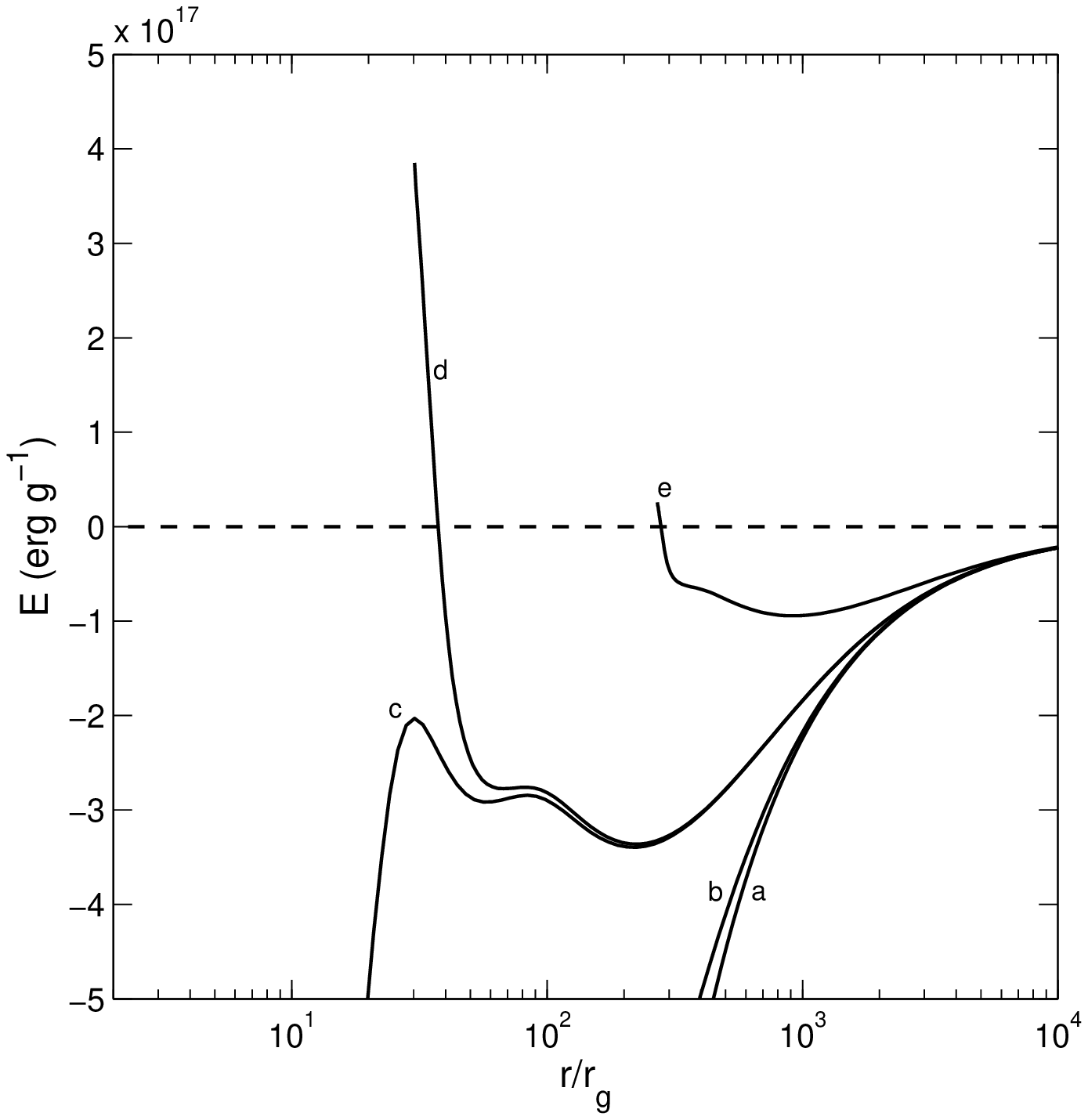} \caption{Specific total energy $E$ as a function of
$r$. Lines $a$, $b$, $c$, $d$ and $e$ are the correspondents of
those in Fig.1, respectively.}\label{fig6}
\end{figure}

\begin{thebibliography}{}
\bibitem[]{} Abramowicz, M. A., Chen, X., Kato, S., Lasota, J.-P., \& Regev, O. 1995, \apj, 438, L37
\bibitem[]{} Abramowicz, M. A., Czerny, B., Lasota, J.-P., \& Szuszkiewicz, E.
1988, \apj, 332, 646
\bibitem[]{} Abramowicz, M. A., Lanza, A., \& Percival, M. J. 1997, \apj, 479, 179
\bibitem[]{} Abramowicz, M. A., Lasota, J.-P., \& Igumenshchev, I.
V. 2000, \mnras, 314, 775
\bibitem[]{} Abramowicz, M. A., Lasota, J.-P., \& Xu, C. 1986, in IAU Symp. 119, Quasars, ed. G. Swarup \& V. K. Kapahi (Dordrecht: Reidel), 371
\bibitem[]{} Artemova, Y. V., Bisnovatyi-Kpgan, G. S., Igumenshchev,
I. V., \& Novikov, I. D. 2006, \apj, 637, 968
\bibitem[]{} Chen, L.-H., \& Wang, J.-M. 2004, \apj, 614, 101
\bibitem[]{} Chen, X., Abramowicz, M. A., Lasota, J.-P.,
Narayan, R., \& Yi, I. 1995, \apj, 443, L61
\bibitem[]{} Chen, X., \& Taam, R. E. 1993, \apj, 412, 254
\bibitem[]{} Frank, J., King, A., \& Raine, D. 2002, Accretion Power in Astrophysics (Cambridge: Cambridge Univ. Press)
\bibitem[]{} Gierli\'nski, M., \& Done, C. 2004, \mnras, 347, 885
\bibitem[]{} Gu, W.-M., \& Lu, J.-F. 2007, \apj, 660, 541 (GL07)
\bibitem[]{} Gu, W.-M., Xue, L., Liu, T., \& Lu, J.-F. 2008, in preparation
\bibitem[]{} H\={o}shi, R. 1977, Prog. Theor. Phys., 58, 1191
\bibitem[]{} Janiuk, A., Czerny, B., \& Siemiginowska, A. 2002, \apj, 576, 908
\bibitem[]{} Kato, S., Fukue, J., \& Mineshige, S. 1998, Black-Hole Accretion
Disks (Kyoto: Kyoto Univ. Press)
\bibitem[]{} Narayan, R., \& Yi, I. 1994, \apj, 428, L13
\bibitem[]{} Narayan, R., \& Yi, I. 1995, \apj, 444, 231
\bibitem[]{} Paczy\'{n}ski, B., \& Wiita, P. J. 1980, \aap, 88, 23
\bibitem[]{} Shakura, N. I., \&
Sunyaev, R. A. 1973, \aap, 24, 337
\bibitem[]{}  Watarai, K. 2006, \apj, 648, 523
\bibitem[]{}  Watarai, K., Ohsuga, K., Takahashi, R., \& Fukue, J.
2005, \pasj, 57, 513
\end{thebibliography}
\end{document}